\documentclass[12pt]{article}
\advance\topmargin by -1.6cm

\newenvironment{eq}
{\[\begin{array}}{\end{array}\]{}}

\let\rvec=\vec        

\def\ind{\stackrel{~\rm ind}\fune}

\def\lq{\hskip-0.5mm\setminus\hskip-0.5mm} 
  
 \def\({\Bigl(}
\def\){\Bigr)}   
 \def\|{\Big|}
\def\then{~\Rightarrow~}  
 \def\o{\circ}
    
\def\x{\times}
   
\def\ox{\otimes}

\def\PL{\displaystyle \bigoplus}

\def\BIGUPLUS{\displaystyle \biguplus}
\def\mid{\big\bracevert}

\def\sub{\subseteq}
\def\subnoteq{\subset}

\def\supnoteq{\supset}
\def\and{\wedge}

\def\rin{{\,\in\kern-.42em\in}}

\def\spec{\,{\rm spec}\,}
\def\rep{\,{\rm rep}\,}

\def\tr{{\,{\rm tr }\,}}
\def\det{\,{\rm det }\,}
\def\id{\,{\rm id}}

\def\centr{\,{\rm centr}\,}
\def\Kern{\,{\rm kern}\,}

\def\card{\ro{\,card\,}}

\def\xs{\lvec \x}

\def\rep{{{\bf rep\,}}}

\def\A{{\,{\rm A\kern-.55emA}}}
\def\B{{\,{\rm I\kern-.2emB}}}
\def\C{{\,{\rm I\kern-.55emC}}}
\def\E{{\,{\rm I\kern-.2emE}}}
\def\G{{\,{\rm I\kern-.55emG}}}
\def\H{{{\rm I\kern-.2emH}}}
\def\I{{\,{\rm I\kern-.2emI}}}
\def\K{{\,{\rm I\kern-.2emK}}}
\def\L{{\,{\rm I\kern-.2emL}}}
\def\M{{\,{\rm I\kern-.16emM}}}
\def\N{{\,{\rm I\kern-.16emN}}}
\def\Q{{\,{\rm I\kern-.5emQ}}}
\def\R{{{\rm I\kern-.2emR}}}
\def\S{{\,{\rm I\kern-.42emS}}}
\def\T{{\,{\rm I\kern-.37emT}}}
\def\UU{{\,{\rm I\kern-.51emU}}}
\def\Z{{\,{\rm Z\kern-.32emZ}}}

\def\al{\alpha}  \def\be{\beta} \def\ga{\gamma}
    
    \def\io{\iota}
   \def\la{\lambda}   \def\si{\sigma}
   \def\om{\omega} 
\def\phi{\varphi} 
   
    \def\La{\Lambda}

\def\vec#1{\underline{\bf vec}_{#1}}

\def\GL{{\bf GL}}  
\def\SL{{\bf SL}}
\def\U{{\bf U}} 
\def \UL{{\bf UL}} 
   
\def\SU{{\bf SU}} 
 
\def\SO{{\bf SO}}

 \def\D{{\bl D}}

\def\norm#1{\parallel #1\parallel}
\def\d#1{{\check{#1}}}

\def\brack#1{\lbrack#1\rbrack}

\def\ro#1{{\rm #1}}
\def\bl#1{{\bf {#1}}}
\def\cl#1{{\cal #1}}

\def\lvec#1{\stackrel{\leftarrow}{#1}}

\def\dprod#1#2{\langle#1,#2\rangle}

\def\map{\longrightarrow}
\def\inmap{\hookrightarrow}

\def\dmap{\Big\downarrow}

\def\mape{\longmapsto}

\def\fune{~\rule[-.1mm]{.5mm}{2mm}\rule[.75mm]{9mm}{.5mm}\hskip-3.5mm>}

\def\Diagr#1#2#3#4#5#6#7#8{\matrix{\noalign{\vskip5mm}
      &              &{\scriptstyle #5}&              &     \cr
      & #1           & \map           & #2           &     \cr
{\scriptstyle #8}   &\dmap         &    &\dmap  &{\scriptstyle#6} \cr
      & #4           & \map           & #3           &     \cr
      &              &{\scriptstyle#7}&              &     \cr
\noalign{\vskip5mm}             }}

\begin{document}
\begin{titlepage} 
\hfill MPI-PhT/01-05 

\hfill hep-th/

\vskip25mm
\centerline{\bf  SPACETIME AS THE MANIFOLD OF}
\vskip5mm 
\centerline{\bf THE INTERNAL SYMMETRY ORBITS} 
\vskip5mm 
\centerline{\bf  IN THE EXTERNAL SYMMETRIES}
\vskip20mm 
\centerline{Heinrich Saller\footnote{\scriptsize 
hns@mppmu.mpg.de}  
}
\centerline{Max-Planck-Institut f\"ur Physik and Astrophysik}
\centerline{Werner-Heisenberg-Institut f\"ur Physik}
\centerline{M\"unchen}
\vskip25mm

\centerline{\bf Abstract}
\vskip7mm

Interactions and particles in the standard model 
are characterized by the action
of internal and external symmetry groups.
The four symmetry regimes involved  
are related to each other in the context of induced group representations. 
In addition to Wigner's induced
representations of external Poincar\'e group operations, parametrized by
energy-momenta,
and the induced internal hyperisospin representations,
  parametrized by the
 standard model  Higgs field, the external operations, 
including the Lorentz group,
can be  considered to be induced also by representations of the internal
hypercharge-isospin group.
In such an interpretation nonlinear spacetime  
is parametrized by the orbits of the internal action group in the
external action group.

\vskip5mm

\end{titlepage}

{\small \tableofcontents}

\newpage

\section{Introduction}

Particle physics as described in the standard model for electroweak and strong
interactions is characterized by four symmetry regimes.
First one has the  external spacetime related transformation
groups, the Lorentz group $\SO_0(1,3)$ with its double cover $\SL(\C^2)$,
 and the internal compact groups - $\U(1)$ for hypercharge, $\SU(2)$ for 
 isospin and
 $\SU(3)$ for color acting on the quantum fields 
 which describe the interaction.
From these symmetries for the interaction, one has to distinguish sharply
the external and internal symmetries for the asymptotic particle states.
 A particle is characterized by
 one translation eigenvalue, its mass, and one space rotation number, its spin
 for nontrivial mass and its polarization for the massless case. The rotation
  invariants are related 
 to external subgroups - spin $\SU(2)$ and 
 polarization  $\SO(2)$.
  With respect to the internal symmetry only an abelian electromagnetic
  $\U(1)$ remains
  as symmetry for the particles. Color symmetry is confined and 
  hypercharge-isospin is spontaneously broken.
  The  groups involved
  
\begin{eq}{c}

\begin{array}{|c||c|c|}\hline
 &\hbox{internal}&\hbox{external}\cr\hline\hline
\hbox{interactions}&\U(1)\o[\SU(2)\x\SU(3)] &
\SL(\C^2)
\cr\hline  
\hbox{particles}&\U(1) &\SU(2),~\SO(2)\cr\hline  
\end{array}\cr\cr
\hbox{four symmetry regimes}  
\end{eq}show subgroup relations in the vertical direction,
i.e. from
particles to interactions \cite{S003}. 
It will be argued below that there is also a horizontal
subgroup relation involved, i.e. from internal to external
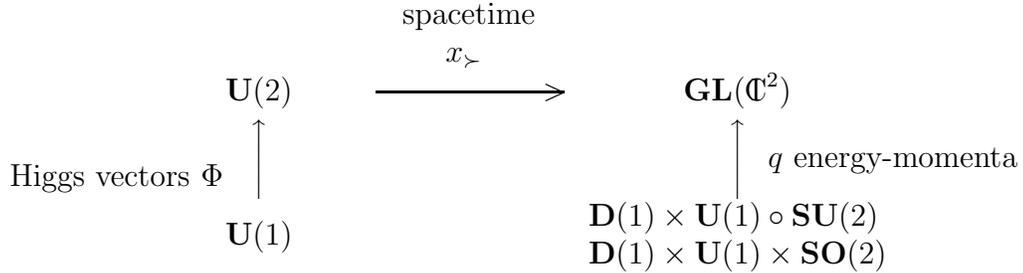
\begin{eq}{l}  
\begin{array}{rcccl}
&&\begin{array}{c}
\hskip8mm\hbox{ spacetime}\cr
\hskip8mm x_\succ\end{array}&&\cr
&\U(2)&\hskip-20mm\rule[1mm]{25mm}{0.2mm}\hskip-4mm>\hskip-30mm&\GL(\C^2)&\cr
\noalign{\vskip1mm}
\begin{array}{r}
\cr
\hbox{Higgs vectors }\Phi\end{array}\hskip-5mm&\Bigg\uparrow
&&\Bigg\uparrow &\hskip-21mm q\hbox{ energy-momenta}\cr
&\U(1)& &\begin{array}{l}
\D(1)\x\U(1)\o\SU(2)\cr
\D(1)\x\U(1)\x\SO(2)\end{array}&\cr
\end{array}  
\end{eq}In this
diagram the external interaction group (upper right) is the Lorentz 
covering group $\SL(\C^2)$, 
supplemented with a
dilatation group $\D(1)=\exp\R$ (causal group) and a 
 phase group $\U(1)=\exp i\R$ 
 (fermion number group), i.e. the full linear group $\GL(\C^2)$.   
A corresponding extension 
with $\GL(\C)=\U(1)\x\D(1)$ is used for the external particle groups
(lower right).
For the internal  interaction symmetry (upper left), 
the color group $\SU(3)$ is omitted.
The three arrows for  inclusion relations
will be related below to induced  
representations. They are labeled with manifold parameters,
the Higgs parameters $\Phi\in\C^2$ with $\norm \Phi^2=M^2>0$  
for the internal induction from particle symmetry (lower left) to
interaction symmetry,
the mass shell energy-momenta $q\in\R^4$ with $q^2=m^2\ge0$ for   
the corresponding external induction, and 
strictly future spacetime parameters $x_\succ\in\R^4$ with 
$x_\succ^2>0$ and $x_{_\succ0}>0$ 
parametrizing the induction from internal to external interaction symmetries.   
To motivate and to   understand  these 
transmutations from groups to supgroups is the aim of this paper.

  \section
  [External Transmutation from Particles to Interactions]
  {External Transmutation\\from Particles to Interactions}
  
According to Wigner \cite{WIG} particles are embedded into
an irreducible definite unitary action of the Poincar\'e group
$\R^4\xs \SO_0(1,3)$ as the semidirect product of the
orthochronous Lorentz group $\SO_0(1,3)$
with the spacetime translations $\R^4$. 	
The 
infinite dimensional Poincar\'e group re\-pre\-sen\-ta\-tions 
are induced \cite{MACK,FOL} by
finite dimensional irreducible re\-pre\-sen\-ta\-tions of direct product subgroups
where the homogeneous factor comes from
energy-momentum  fixgroups (`little groups'):
The rest system rotations $\SO(3)$ for energy-momenta 
$q\in\R^4$ with $q^2>0$,
the noncompact group $\SO_0(1,2)$ for energy-momenta with  $q^2<0$
and the axial rotations $\SO(2)$ around the momentum direction 
as `fixgroup in the fixgroup' 
$\R^2\xs \SO(2)\subnoteq \SO_0(1,3)$ for nontrivial 
energy-momenta with $q^2=0$.
Only particles 
with the re\-pre\-sen\-ta\-tions for causal momenta $q^2\ge 0$   
and compact little groups $\SO(3)$ and $\SO(2)$ are found. 

With respect to the halfinteger spin particles the twofold covering
simply connected groups $\SL(\C^2)\supnoteq \SU(2)$ for
$\SO_0(1,3)\supnoteq\SO(3)$ are used
\begin{eq}{l}
\begin{array}{rllll}
u\in\SU(2)\then&O(u)^a_b&={1\over 2}\tr u\si^a u^*\si^b&\in\SO(3)&\cong
\SU(2)/\I(2)\cr
s\in\SL(\C^2)\then&
\La(s)^i_j&={1\over 2}\tr s\si^i s^*\d\si_j&\in\SO_0(1,3)
&\cong\SL(\C^2)/\I(2)\end{array}\cr
\hbox{with } \centr\SL(\C^2)=\centr\SU(2)=\I(2)=\{\pm\bl1_2\}
\end{eq}The traces involve  the hermitian 
\begin{eq}{l}
\hbox{Pauli-Weyl matrices: }\si^i=(\bl 1_2,\si^a)=\d\si_i,~~
\left\{\begin{array}{rl}
a&=1,2,3\cr
i&=0,1,2,3\cr\end{array}\right.\cr
\end{eq}Therewith the twofold   cover 
of the Poincar\'e group comes with the multiplication law 
\begin{eq}{rl}
\R^4\xs\SL(\C^2)\ni (x,s)
\hbox{ with }&\left\{
\begin{array}{l}
x=x_j\si^j=
{\scriptsize\pmatrix{x_0+x_3&x_1-ix_2\cr x_1+ix_2&x_0-x_3\cr}}\in\R^4\cr
(x,s)\o(x',s')=(x+s\o x'\o s^*, s\o s')\cr\end{array}\right.\cr
\end{eq}

The induction procedure used for massive and massless 
particles is symbolized 
with the re\-pre\-sen\-ta\-tion equivalence  classes $\rep$ as follows
\begin{eq}{l}
\rep[\R\x \SU(2)]\uplus\rep[\R\x \SO(2)]\ind
\rep[\R^4\xs\SL(\C^2)] 
\end{eq}The discrete invariants $2J\in\N$ for $\SU(2)$ and 
$\pm z\in\Z$ for $\SO(2)$ give spin and polarization resp., the continuous 
invariant $q^2=m^2\ge 0$ for the  translations
gives the corresponding mass.
According to Wigner's particle definition, confined quarks are no particles.

In the case of spin $\SU(2)$, 
the transition from a massive particle rest system, defining a time direction,
 to the Lorentz group 
action compatible framework is performed with the boost re\-pre\-sen\-ta\-tions,
parametrized by the three real numbers in the energy-momenta ${q\over m}$
\begin{eq}{l}
s({q\over m})= e^{{\rvec\be\rvec\si\over 2}}
= {\sqrt{m+q_0\over 2m}\brack{\bl1_2+{\rvec q\rvec\si\over m+q_0}}}
\in\SL(\C^2) \hbox{ with }\left\{\begin{array}{l}
\rvec\be={\rvec q\over|\rvec q|}\ro{artanh}{|\rvec q|\over q_0}\cr
q_0=\sqrt{m^2+\rvec q^2}\cr
s({q\over m}) m\bl1_2 s^*({q\over m})=q_j\si^j\cr\end{array}\right.
\end{eq}In the case of polarization  $\SO(2)$, 
the transition from a space  system with the 
distinguished
polarization axis as 3rd direction 
 to a rotation group 
action compatible framework is performed with the 
2-sphere re\-pre\-sen\-ta\-tions, parametrized by the two real numbers in
the  momenta ${\rvec q\over |\rvec q|}$
\begin{eq}{l}
 u({\rvec q\over |\rvec q|})
=e^{i{\rvec\al\rvec\si\over 2}}=
 \sqrt{{|\rvec q|+q^3\over 2|\rvec q|}}
\brack{\bl 1_2+i{\rvec q_\perp\rvec\si\over |\rvec q|+q^3}}\in\SU(2)
\hbox{ with }\left\{\begin{array}{l}
\rvec\al=
{\rvec q_\perp\over|\rvec q_\perp|}\arctan{|\rvec q_\perp|\over |\rvec q|}\cr
\rvec q_\perp=(q_2,-q_1,0)\cr
u({\rvec q\over |\rvec q|})|\rvec q|\si^3u^*({\rvec q\over |\rvec q|})
=\rvec q \rvec\si \end{array}\right.
\end{eq}Such linear re\-pre\-sen\-ta\-tions of coset representatives, here
$s({q\over m})$ and $\hat s({q\over m})=s^{-1*}({q\over m})$ for
the boosts  $\SL(\C^2)/\SU(2)\cong\SO_0(1,3)/\SO(3)$ in the
two fundamental Weyl re\-pre\-sen\-ta\-tions (often introduced as solutions of the
Dirac equation)
and 
$u({\rvec q\over |\rvec q|})$ for 
the 2-sphere $\SU(2)/\SO(2)\cong\SO(3)/\SO(2)$ in the 
fundamental Pauli re\-pre\-sen\-ta\-tion,
will be called transmutators. They have 
a characteristic hybrid transformation
property: The left action with the supgroup
gives the transmutator for the
transformed momenta up to a right action with the subgroup
\begin{eq}{llll}
\la\in\SL(\C^2)&\then \la\o s({q\over m})=s(\La(\la).{q\over m})\o u&\hbox{with }
u=u({q\over m},\la)&\in\SU(2)\cr
v\in\SU(2)&\then v\o u({\rvec q \over |\rvec q|})
=u(O(v).{\rvec q \over |\rvec q|})\o a&\hbox{with }
a=a({\rvec q \over |\rvec q|},v)&\in\SO(2)\cr
\end{eq}

External transmutators show up in the harmonic
(Fourier) analysis of quantum fields with respect to the 
particle-antiparticle $(\ro u,\ro a)$ creation and annihilation
operators involved, e.g. for the left and right handed
Weyl component of a Dirac electron  field 
\begin{eq}{l}
\bl\Psi(x)=
{\scriptsize\pmatrix{\bl r^{\dot A}\cr\bl l^{ A}\cr}}(x)
= \int {d^3 q\over (2\pi)^3}
{\scriptsize\pmatrix{
s({q\over m})^{\dot A}_C&
{ e^{xiq}\ro u^C(\rvec q)+ e^{-xiq}\ro a^{\star C}(\rvec q)\over\sqrt2}
 \cr
\hat s({q\over m})_C^{ A}&{ e^{xiq}\ro u^C(\rvec q)- e^{-xiq}\ro a^{\star C}(\rvec q)\over\sqrt2}
\cr}}\cr
\end{eq}The infinite dimensionality
($\R^3$-cardinality) of the definite unitary 
representations of the noncompact Poincar\'e group is 
seen in the momentum integral 
$\int {d^3 q\over (2\pi)^3}\cong{\PL_{\rvec q\in\R^3}}$
over all transmutators.

Higher spin and polarization fields,
 e.g. the massive weak vector bosons or
the massless electromagnetic vector potential, need transmutators which are 
 products of the two fundamental 
Weyl transmutators and the fundamental  Pauli transmutator resp., e.g.
\begin{eq}{rll}
\La({q\over m})^i_j&={1\over 2}\tr s({q\over m})
\si^i s^*({q\over m})\d\si_j&\in\SO_0(1,3)\cr
O({\rvec q\over |\rvec q|})^a_b&
={1\over 2}
\tr u({\rvec q\over |\rvec q|})\si^a u^*({\rvec q\over |\rvec q|})\si^b
&\in\SO(3)\cr
\end{eq}

  \section
  [Internal Transmutation from Particles to Interactions]
  {Internal Transmutation\\from Particles to Interactions}

In addition to the external 
rotation and translation properties particles are characterized 
also by  particle-antiparticle $\U(1)$-symmetries, e.g.
the electromagnetic charge number or a
fermion-antifermion number, e.g. for the neutrinos or the neutron.
In the standard model of electroweak interactions the electromagnetic 
real 1-dimensional abelian internal $\U(1)$-symmetry is the
only remaining symmetry from the  real 12-dimensional 
rank 4 hyperisospin-color group.
Particles have no isospin or color symmetry.
E.g. the proton-neutron dublet displays  the isospin multiplicity
two, but - with
the different masses -  no isospin $\SU(2)$-symmetry.

In the standard model,
the electromagnetic group $\U(1)$
 is the only proper  fixgroup ('little group')
for the  hyperisospin group $\U(2)$
acting on the  complex 2-dimensional Hilbert space with the   Higgs field
 $\Phi\in\C^2$ with nontrivial scalar product
$\norm\Phi^2=M^2>0$. The internal induction from
electromagnetic $\U(1)$ to hyperisospin $\U(2)$
\begin{eq}{l}
\rep\U(1)\ind\rep \U(2)
\end{eq}is in analogy to the external inductions.
The 
 analogy to the rest systems, defined by $q_j\si^j=m\bl1_2$ up to 
 rotations $\SO(3)$,
and the polarization systems, defined by $\rvec q\rvec \si=|\rvec q|\si^3$
up to axial rotations $\SO(2)$,
is  the electromagnetic system which is 
defined  by 
${\scriptsize\pmatrix{\Phi_1\cr \Phi_2\cr}}
={\scriptsize\pmatrix{0\cr M\cr}}$
up to electromagnetic
transformations $e^{i(\bl1_2+\tau^3)\ga_0}\in \U(1)_+$.
The internal induction
 employs the Higgs field defined transformation 
\begin{eq}{l}
v({\Phi\over M})
={1\over M}{\scriptsize\pmatrix{\Phi^{\star 2}&\Phi_1\cr
-\Phi^{\star 1}&\Phi_2\cr}}\in\U(2)
\hbox{ with }\left\{\begin{array}{l}
\norm \Phi^2=M^2>0\cr
v({\Phi\over M}){\scriptsize\pmatrix{0\cr M\cr}}=
{\scriptsize\pmatrix{\Phi_1\cr \Phi_2\cr}}
\end{array} \right.\cr
\end{eq}from the electromagnetic system
to the hyperisospin $\U(2)$ compatible framework. 
The Goldstone manifold $\U(2)/\U(1)_+$ 
involved is parametrized with the three real parameters in ${\Phi\over M}$.
The hybrid transformation looks like 
\begin{eq}{l}
u\in\U(2)\then u\o v({\Phi\over M})
=v(u.{\Phi\over M})
\o t \hbox{ with }
t=t(u)\in\U(1)_+\cr
\end{eq}

The transition from the interaction parametrizing fields with
hyperisospin symmetry to the 
particle electromagnetic  symmetry is performed
with the Higgs transmutator
$v({\Phi\over M})$ (in analogy to the Weyl and 
Pauli  transmutators),
e.g. from the left-handed lepton isodoublet  $(\bl L_\al )_{\al=1,2}$ 
to the left-handed components for the charged massive lepton field 
and its neutrino which, in turn, are transmuted to their
particle systems as described in the last section
\begin{eq}{l}
\bl L^{ A}_\al=v({\Phi\over M})_\al^.
{\scriptsize\pmatrix{\nu^{ A}\cr \bl l^{ A}\cr}},~~
\left\{\begin{array}{rl}

\nu^{ A}(x)&=\dots\cr
\bl l^{ A}(x)&=\dots\end{array}\right.
\end{eq}In contrast to the external case 
 only compact groups are involved.
Their irreducible representations are finite dimensional. Therefore
there is no analogue to the momentum integral, necessary  for 
the infinite dimensional representation of the  external
noncompact groups. 

Higher isospin fields, e.g. the isotriplet 
gauge  field, need transmutators which are  products of 
the fundamental
Higgs transmutator, e.g.
\begin{eq}{l}
O({\Phi\over M})^a_b
={1\over 2}
\tr v({\Phi\over M})\tau^a v^*({\Phi\over M})\tau^b\in\SO(3)
\end{eq}

\section
[The Operational  Triunit: Internal-Spacetime-External] 
{The Operational  Triunit:\\Internal-Spacetime-External} 

The transition from the large  operational
symmetry group  of the standard model interactions to the 
small symmetry groups  of the 
related particles involves the external Weyl-Pauli transmutations
and the internal Higgs transmutation
\begin{eq}{r}
\rep[\U(1)\x \R\x \SU(2)]\uplus\rep[\U(1)\x\R\x \SO(2)]\cr\ind
\rep[\U(2)\x[\R^4\xs\SL(\C^2)]] 
\end{eq}

If $\SU(3)$ color fields are included the right hand side
has to be written with the hyperisospin-color group
 \cite{HUCK,S921}
whose three  factors are correlated via the centrum 
$\I(2)\x\I(3)=\I(6)=\{z\in\Z\mid z^6=1\}$ of the nonabelian
factor
\begin{eq}{l}
\rep\U(1)\ind\rep \U(2\x3)\hbox{ with }
\U(2\x 3)={\U(1)\x\SU(2)\x\SU(3)\over \I(2)\x\I(3)}
\end{eq}For the following considerations the color group
is excluded. It cannot 
be described in the structures below,
its occurence has to be explained differently, 
e.g. as proposed in  \cite{S982,S002}.

The three factors in the standard model interaction symmetry 
$\U(2)\x[\R^4\xs\SL(\C^2)]$
describe the  internal 
operations, the  spacetime translations and the homogeneous
external operations resp. 
Such a product   constitutes 
a characteristic structure \cite{MACK,FOL,FULHAR} 
occuring for re\-pre\-sen\-ta\-tions
of a group $G$ induced by re\-pre\-sen\-ta\-tions of a subgroup $U\sub G$.
In the re\-pre\-sen\-ta\-tion induction, which will be described in more detail below,
 the group $G$ is decomposed
into disjoint subgroup $U$-orbits 
and representatives  $(U\lq G)_{\rm repr}$ for the cosets $U\lq G$
\begin{eq}{l}
G=U\x (U\lq G)_{\rm repr}={\BIGUPLUS_{{\rm repr }~ k_r\in G}}Uk_r
\end{eq}For notational convenience the left classes $Uk$,
i.e. the $U$-orbits under left multiplication are taken
\begin{eq}{l}
u\in U:~~L_u: G\map G,~~L_u(k)=uk
\end{eq}

To establish  the standard model operations as an
example for the abstract structure
\begin{eq}{l}
\U(2)\x[\R^4\xs\SL(\C^2)]~\stackrel ?\sim~ 
U\x [(U\lq G)_{\rm repr}\xs  G]
\end{eq}the Lorentz group cover $\SL(\C^2)$  is filled up
by a   phase  $\U(1)$-group (fermion number) and a
 dilatation group $\D(1)$ (causal group)
to the full linear group $\GL(\C^2)$, a real 8-dimensional Lie group
\begin{eq}{l}
\GL(\C^2)=\D(\bl1_2)\x \UL(2)
\end{eq}The direct unimodular factor involved is the
centrally correlated product of two normal subgroups,
the fermion number  and the Lorentz covering group
\begin{eq}{l}
\UL(2)=\U(\bl 1_2)\o\SL(\C^2)=\{ g\in\GL(\C^2)\mid |\det g|=1\}\cr 
\U(\bl 1_2)\cap\SL(\C^2)=\I(2)=\{\pm \bl1_2\},~~\left\{\begin{array}{rl}
\UL(2)/\SL(\C^2)&\cong\U(1)\cr
\UL(2)/\U(\bl1_2)&\cong\SL(\C^2)/\I(2)\cr
&\cong\SO_0(1,3)\end{array}\right. 
\end{eq}Therewith the triad $U\x [(U\lq G)_{\rm repr}\xs  G]$
 of
the internal-spacetime-external transformations
will be defined with a maximal compact subgroup $\U(2)$, defining
the  internal operations,  in the full group $\GL(\C^2)$, defining
the external operations   
\begin{eq}{l}
\hbox{\bf operational triunit: }\U(2)\x[\D(2)\xs\GL(\C^2)]
\end{eq}The manifold 
of hyperisospin $\U(2)$ orbits in the full external group $\GL(\C^2)$
is a real 4-dimensional rank 2  symmetric space
$\D(2)$ which will be  used as model for nonlinear
spacetime \cite {S97,S982}. It  has as representatives  
the hermitian invertible $2\x2$-matrices which
can also be parametrized by the translations of  
the strictly  future lightcone 
\begin{eq}{rl}
(\U(2)\lq \GL(\C^2))_{\rm repr}=\D(2)
=\{k\in\GL(\C^2)\mid& k=k^*
={\scriptsize\pmatrix{k_0+k_3&k_1-ik_2\cr k_1+ik_2&k_0-k_3\cr}}\cr
&\hbox{ and }\spec k >0\}\cr
\end{eq}All $2\x2$-matrices with $\U(2)$-conjugation
constitute a $C^*$-algebra with the natural
spectral order and the  polar decomposition
of the full group into internal compact operations and noncompact spacetime
\begin{eq}{l}
\GL(\C^2)=\U(2)\x\D(2),~~k=u\o |k|,~~|k|=\sqrt{k^*\o k}
\end{eq}

In the general structure,
 the  group $G$ acts
on the left orbits $Uk$ of a subgroup $U$ by right inverse multiplication
which may look quite complicated for the chosen orbit representatives 
\begin{eq}{rcccl} 
g\in G:~~R_g:& U\lq G&\map& U\lq G,&R_g(Uk)=Ukg^{-1}\cr
 &(U\lq G)_{\rm repr}&\map& (U\lq G)_{\rm repr},&k_r\mape k_{r'}
\hbox{ for }k_rg^{-1}=u k_{r'}\cr
&&&&\hbox{with }u=u(k_r,g^{-1})\in U\cr
\end{eq}In the physical structure proposed  one obtains the action of the
full external group $\GL(\C^2)$ on the nonlinear spacetime $\D(2)$
\begin{eq}{rl}
g\in \GL(\C^2):~~\D(2)\map \D(2),&  
|k|\mape |k'| \hbox{ for }|k|\o g^{-1}=u\o|k'|\cr
&\hbox{with }u=u(|k|,g^{-1})
\in\U(2)\cr
\then &|k'|=\sqrt{g^{-1*}\o |k|^2\o g^{-1}}=|k\o g^{-1}|
\end{eq}

The tangent space of the symmetric space $\D(2)$ constitutes the
spacetime translations with the  faithful
action of the causal Lorentz group
\begin{eq}{l}
\log\D(2)=\{x=x_j\si^j\mid e^x=|k|\in\D(2)\}\cong\R^4\cr
g=e^{{\be_0+i\al_0\over 2}}s\in\GL(\C^2):~~x\mape g\o x\o g^*\cr
\then
{1\over 2}\tr g\si^i g^*\d\si_j
=e^{\be_0}\La(s)\in\D(1)\x\SO_0(1,3)\cong\GL(\C^2)/\U(\bl1_2)\cr
\end{eq}

\section{Internal-External Actions on Standard Model Fields}

The transformation behavior of  fields
with respect to external 
Lorentz and internal hyperisospin operations is quite different:
The fields used in the standard model
with the operations $\U(2)\x[\R^4\xs\SL(\C^2)]$ , e.g.
the left-handed lepton isodoublet $\{\bl L_\al^A\}_{\al=1,2}^{A=1,2}$, 
 map
the spacetime translations $\R^4$ into a complex vector
space 
\begin{eq}{l}
\bl L_\al^A:\R^4\map W\ox V^T,~~x\mape \bl L_\al^A(x) 
\end{eq}The value space is  the tensor product of
a finite dimensional space $W$ with the re\-pre\-sen\-ta\-tion of 
 hyperisospin $\U(2)$, in the lepton 
 isodoublet example  the defining re\-pre\-sen\-ta\-tion 
on $W\cong\C^2$ with $\U(1)$-hypercharge number  $y=-{1\over2}$
 \begin{eq}{rll}
 D:\U(2)&\map\GL(W),& D(u)=u=e^{-i\ga_0\bl1_2+i\rvec\ga\rvec\tau\over 2}\cr 
 \U(2)\x W&\map W,&u.\bl L_\al^A=u_\al^\be
\bl L_\be^A\cr
\end{eq}and another finite dimensional vector space $V$ 
with
a Lorentz group re\-pre\-sen\-ta\-tion, in the example the 
defining left handed  Weyl 
re\-pre\-sen\-ta\-tion on $V\cong\C^2$ 
 \begin{eq}{l}
T: \SL(\C^2)\map\GL(V),~~ T(s)=s= e^{(i\rvec \al+\rvec\be){\rvec\si\over 2}}\cr 
\end{eq}Since also the
spacetime translations $\R^4$
are acted upon with the
Lorentz group,
the field as a mapping between two vector spaces
with Lorentz group action 
transforms $\bl L\mape \bl L_s$ as given by the commutative diagram \cite{LIE13}
\begin{eq}{l}
\Diagr{\R^4}{\R^4}{V^T}{V^T}{\La(s)}{\bl L_s}s{\bl L},~~\begin{array}{l}
\La=\La(s)\in\SO_0(1,3)\cr
\bl L_s(\La.x)=s.\bl L(x)\cr\end{array}\cr
\SL(\C^2)\x V^T\map V^T,~~ (\bl L_s)_\al^A(x)=  \bl L^B_\al(\La^{-1}.x)s^A_B
\end{eq}For notational convenience the dual space $V^T$ (linear $V$-forms)
is used.

Both internal and external transformation behavior can be collected
into one diagram, e.g. for the lepton 
isodoublet left-handed Weyl field above
\begin{eq}{cc}
\Diagr{\R^4}{\R^4}{W\ox V^T}{W\ox V^T}
{\La(s)}{\bl L_s}{u\ox s}{\bl L},&
(\bl L_s)_\al^A(x)= u_\al^\be \bl L^B_\be(\La^{-1}.x)s^A_B\cr
\end{eq}or for
the isotriplet gauge vector field $\{\bl A_a^j\}_{a=1,2,3}^{j=0,1,2,3}$
valued in the vector space $W'\ox V'{}^T\cong\C^3\ox\C^4$
\begin{eq}{cc}
\Diagr{\R^4}{\R^4}{W'\ox V'{}^T}{W'\ox {V'}^T}
{\La(s)}{\bl A_s}{O(u)\ox \La(s)}{\bl A},&\begin{array}{c}
(\bl A_s)_a^j(x)= O_a^b \bl A^k_b(\La^{-1}.x)\La^j_k\cr
O=O(u)\in\SO(3)\cr\end{array}\cr
\end{eq}

These transformation properties are compared in the next sections with
the transformation properties occuring for induced re\-pre\-sen\-ta\-tions.

\section{Induced Representations}

The structure of induced re\-pre\-sen\-ta\-tions as used e.g. for
Wigner's particles classification can be sketched 
for our purposes - without discussion of
topological structures - as follows \cite{MACK,FOL,FULHAR}:

A group $G$-re\-pre\-sen\-ta\-tion induced by the re\-pre\-sen\-ta\-tion
of a subgroup $D:U\map\GL(W)$ on a complex vector space
acts on the subgroup intertwiners, i.e. on the
mappings from the group $G$ into the vector space $W$,
compatible with the action of $U$ on $G$ by left 
multiplication and  on $V$ by the re\-pre\-sen\-ta\-tion $D$
\begin{eq}{l}
\Diagr GGWW{L_u}w {D(u)}w,~~\begin{array}{l}
w(uk)=D(u).w(k)\cr
\hbox{for all }u\in U,~~k\in G\cr
\end{array}  
\end{eq}The intertwiner space dimensionality is the product of the 
$W$-dimensionality with the cardinality of the $U$-orbits, i.e.
in general infinite for Lie groups
\begin{eq}{l}
~\dim_\C W_U(G)=\dim _\C W\cdot\card U\lq G\cr
\end{eq}

The group $G$ action on the vector space with the
intertwiners $w\in W_U(G)$
is defined by the following commutative diagram
which involves the right inverse multiplication $k\mape kg^{-1}$
on the group $G$, not used in the definition of the intertwiners 
\begin{eq}{l}
\Diagr GGWW{R_g}{w_g}{\id_W}w,~~\begin{array}{l}
w_g(k)=w(kg)\cr  
\hbox{for all } k\in G,~~g\in G\cr\end{array}\cr
G\x W_U(G)\map W_U(G),~~w\mape  w_g
\end{eq}Again, both diagrams can be taken together.
With a  decomposition into $U$-orbits and  representatives
$G=U\x (U\lq G)_{\rm repr}={\BIGUPLUS_r}Uk_r$
the induced $G$-re\-pre\-sen\-ta\-tion  reads
\begin{eq}{l}
\Diagr GGWW{L_u\o R_g}{w_g}{D(u)\o \id_W}w,~~\begin{array}{l}
w_g(uk)=D(u). w(kg)\cr
\hbox{for all }u\in U,~k\in G,~~g\in G\cr
w_g(k_r)=D(u). w(k_{r'})\cr
\hbox{for  }k_rg=uk_{r'}\hbox{ with }u=u(k_r,g)\in U\cr
\end{array}\cr  
\end{eq}

\section{Transmutators}

In general, an induced $G$-re\-pre\-sen\-ta\-tion is
infinite dimensional and - in many cases, e.g. for compact groups
 - highly reducible,
e.g. the right regular re\-pre\-sen\-ta\-tion on the algebra
$\C(G)=\{G\map\C\}$ with the group functions, which is
induced by the trivial re\-pre\-sen\-ta\-tion of the trivial subgroup
$U=\{e\}$ on the numbers $\C$, or 
the $G$-representation on an intertwiner space  $W_U(G)$.

The  group functions $\C(G)$ contain  - up to isomorphy -
the representation space of  each finite dimensional $G$-re\-pre\-sen\-ta\-tion
\begin{eq}{rl}
T:G\map\GL(V)
\end{eq} via the representation  matrix elements,
isomorphic to $V\ox V^T$
\begin{eq}{l}
T(g):V\map V\cr
V\ox V^T\cong \{T_\om^v\mid 
v\in V,\om\in V^T\}\subnoteq\C(G)\hbox{ with }\left\{\begin{array}{l}
T^v_\om:  G\map \C\cr
T^v_\om(k)=\dprod\om{T(k).v}\cr
T^v_\om(kg)=T^{T(g).v}_\om(k)\end{array}\right.
\end{eq}

A decomposition of a $G$-re\-pre\-sen\-ta\-tion  into
$U$-re\-pre\-sen\-ta\-tions with  projectors $\{\cl P_\io\}_\io$
\begin{eq}{l}
V={\PL_\io}W_\io,~~T[U].W_\io\sub W_\io,~~T|_U=D={\PL_\io}D_\io\cr
\cl P_\io:V\map W_\io,~~D_\io:U\map\GL(W_\io),~~
D_\io(u):W_\io\map W_\io
\end{eq}and an orbit decomposition of the full group
$G=U\x (U\lq G)_{\rm repr}={\BIGUPLUS_r}Uk_r$
give rise to transmutators which are valued in
the tensors $W_\io\ox V^T$ 
as products of  the $G$-space $V$ and a $U$-subspace $W_\io$ 
\begin{eq}{l}
T_\io:G\map W_\io\ox V^T,~~T_\io(uk_r)=D_\io(u)\o\cl P_j\o T(k_r):V\map W_\io
\end{eq}If $V\cong \C^n$, then $T(k)$ has an $n\x n$-matrix form. 
If $W_\io\cong\C^m$ with $m\le
n$, then $D_\io(u)$ has an $m\x m$-matrix form 
and  $T_\io(k_r)$  an $m\x n$-matrix form.

All 'right-sided' matrix elements
of a transmutator constitute
a $G$-stable subspace of the  $U$-intertwiners 
\begin{eq}{l}
W_\io\ox V^T\cong\{T_\io^v\mid v\in V\}\subnoteq W_{\io, U}(G)\hbox{ with }\left\{
\begin{array}{l}
T_\io^v:G\map W_\io\cr
T_\io^v(uk_r)=D_\io(u)\o\cl P_j\o  T(k_r).v\cr
T_\io^v(kg)=T_\io^{T(g).v}(k)\end{array}\right.
\end{eq}

Therewith the intertwiner space $W_U(G)$  contains  -  up to isomorphy -
all tensor products $W\ox V^T$ where $V$ is acted on with
a finite dimensional suprepresentation
of the full group $G$ 
\begin{eq}{l}
D[U].W\sub T[G]. V\then W\ox V^T\inmap W_U(G)
\end{eq}$U\lq G$-transmutators
for irreducible $G$-representations are  building blocks
of induced representations. They
 transform from a vector space $V$
with the action of a group $G$ to a vector subspace $W$ with the action of a
subgroup $U$.
Transmutators with $W=V$ are called complete, 
i.e. all $U$-representations contained in
the $G$-representation are included. Complete transmutators are bijections.

\section{Fields as Internal-External Transmutators}

Spacetime fields $\bl \Psi$ for
the operational triunit $U\x[(U\lq G)_{\rm repr}\xs G]$
will be defined to be transmutators 
from external group $G$-representations on 
a vector space $V$ to internal subgroup
$U$-representations
on a vector subspace $W$. They are  parametrized with the 
orbit manifold $U\lq G$ of 
the possible $U$'s in $G$ 
\begin{eq}{l}
\bl \Psi:(U\lq G)_{\rm repr}\map W\ox V^T,~~\left\{\begin{array}{rll}
U\x W&\map W,&\hbox{(internal)}\cr
G\x V&\map V,&\hbox{(external)}\cr
U\sub G,&W\sub V&\end{array}\right.
\end{eq}The geometrical structure can be formulated also 
in a bundle language.

The internal hyperisospin group $\U(2)$ is a
maximal compact  subgroup of the 
external group $\GL(\C^2)=\D(\bl1_2)\x\UL(2)$ 
with the causal group and the unimodular fermion number-Lorentz  group
cover
$\UL(2)=\U(\bl1_2)\o\SL(\C^2)$ as direct factors. 
Nonlinear spacetime $\D(2)$ parametrizes the 
noncompact manifold $\U(2)\lq\GL(\C^2)$.

\subsection
[The Fundamental  Transmutator on Nonlinear Spacetime]
{The Fundamental  Transmutator\\on Nonlinear Spacetime}

The fundamental  spacetime field for the operational triunit 
\begin{eq}{l}
\U(2)\x[\D(2)\xs\GL(\C^2)]\cr
\end{eq}transmutes from 
the defining  internal $\U(2)$-isodoublet space $W\cong\C^2$
to the defining  external $\SL(\C^2)$-Weyl spinor   space $V\cong\C^2$
\begin{eq}{l}
\bl\Psi_\al^A:\D(2)\map W\ox V^T,~~|k|\mape\bl\Psi_\al^A(|k|)\cr
\end{eq}It has the internal $\U(2)$ and 
the external $\GL(\C^2)$ transformation behavior
\begin{eq}{rll}
\U(2)\x W&\map W,&\bl\Psi_\al^A\mape u_\al^\be\bl\Psi_\be^A\cr
\GL(\C^2)\x V&\map V,&\bl\Psi_\al^A(|k|g)=
\bl\Psi_\al^B(|k|)g_B^A=u(|k|,g)_\al^\be\bl\Psi_\be^A(|k\o g|)\cr
\end{eq}Since  the nonlinear spacetime manifold 
can be parametrized 
as the strictly future lightcone  
$\D(2)\cong \R^4_\succ\subnoteq\R^4$, $|k|=x_\succ$, 
of its  tangent space, the  spacetime translations
$\log\D(2)\cong\R^4$, the fundamental 
isospinor Weyl spinor  field has causal support
without spacelike particle interpretable contributions.
Its spectrum   with respect to the action of the causal group $\D(1)$
has to be investigated to  find its particle interpretable content
which can be defined for all spacetime translations $\R^4$.
First steps on this way have been tried in  \cite{S002}.

The fundamental isospinor-spinor dyad $\{\bl \Psi^A_\al\}_{\al=1,2}^{A=1,2}$ 
for the hyperisospin $\U(2)$ orbits  in the extended Lorentz group 
$\GL(\C^2)$ can be seen 
in some analogy \cite{S982} to the tetrad 
$\{\bl h^\mu_j\}^{\mu=0,1,2,3}_{j=0,1,2,3}$ in general relativity
for  the orbits of the  Lorentz group $\SO_0(1,3)$ in the general 
linear group
$\GL(\R^4)$.

\subsection
[Standard Model Fields as Transmutators
on Linear Spacetime]
{Standard Model Fields as Transmutators\\
on Linear Spacetime}

Without being able so far to determine the spectrum of the causal group
action on the fundamental 
transmutator for a particle interpretation  one may start
less ambitiously and try to interpret the standard model fields
as a linear approximation, i.e. 
as internal-external
transmutators parametrized with  spacetime translations  $\log\D(2)\cong\R^4$
\begin{eq}{l}
\U(2)\x[\R^4\xs\GL(\C^2)]\cr
\end{eq}

Any representation of a group $D:G\map\GL(V)$ is faithful up to its kernel,
a normal $G$-subgroup, i.e. $D[G]\cong G/\Kern D$.
Therefore the representations of the internal hyperisospin group
$\U(2)=\U(\bl1_2)\o\SU(2)$ 
with $\U(\bl1_2)\cap\SU(2)=\I(2)$ have as nontrivial images  three  groups -
the full hyperisospin, the hypercharge and the iso-rotation group
\begin{eq}{rl}
\hbox{$\U(2)$-representation images: }
\U(2),~\U(1)&\cong\U(2)/\SU(2)\cr
\SO(3)&\cong\U(2)/\U(\bl1_2) 
\end{eq}to be compared with the 
three nontrivial representation images of the external unimodular
group, given by the full group, the fermion number and the Lorentz group
\begin{eq}{rl}
\hbox{$\UL(2)$-representation images: }
\UL(2),~\U(1)&\cong\UL(2)/\SL(\C^2)\cr
\SO_0(1,3)&\cong\UL(2)/\U(\bl1_2) 
\end{eq}

There are
  three nontrivial internal-external embeddings -
hyperisospin $\U(2)$ and hypercharge $\U(1)$ into the fermion number-Lorentz
group $\UL(2)$ and  iso-rotations $\SO(3)$ into the Lorentz group $\SO_0(1,3)$
\begin{eq}{l}
\U(2)\inmap\UL(2),~~\U(1)\inmap \UL(2),~~\SO(3)\inmap \SO_0(1,3)
\end{eq}In the standard model the left-handed  Weyl isodoublet field $\bl L$,
the right-handed  Weyl  isosinglet fields $\bl R$
and the  Lorentz vector isosinglet-isotriplet gauge fields $\bl A$
are the corresponding transmutators 
as mappings from the coset tangent space $\log(U\lq G)_{\rm rep}\map W\ox V^T$ 
into an internal-external vector space tensor product
with the faithful action of
the represented images  $D[U]\ox T[G]$ 
\begin{eq}{l}
\begin{array}{rlcl}
\bl L:\R^4&\map \C^2\ox\C^2&\hbox{with}&\U(2)\ox \UL(\C^2)\cr
x&\mape \bl L^A_\al(x),&&\al=1,2;~~A=1,2\cr
\bl R:\R^4&\map \C^2\ox\C^2&\hbox{with}&\U(1)\ox\UL(\C^2)\cr
x&\mape \bl R^{\dot A}_{1,2}(x),&&\hskip19mm\dot A=1,2\cr
\bl A:\R^4&\map \C^4\ox\C^4&\hbox{with}&\SO(3)\x\SO_0(1,3)\cr
x&\mape \bl A^j_{0,a}(x),&&a=1,2,3;~j=0,1,2,3\end{array}\cr

\end{eq}There are two fermionic and one bosonic transmutator. 
With coinciding internal and external representation space
all three transmutators are complete. The right-handed 
2-component Weyl field $\bl R$ 
comprises two isosinglets $\{\bl R_1,\bl R_2\}$, and the 
4-component Lorentz vector field $\bl A$  four internal degrees of freedom, 
an  isosinglet and an isotriplet $\{\bl A_0,\rvec\bl A\}$.

The transition from  those standard fields for the interactions
 to  particles for the state space
  requires 
internal transmutators, parametrized with the Higgs degrees of freedom
(Goldstone manifold), as discussed above
\begin{eq}{rlcl}
v:
 (\U(2)/\U(1)_+)_{\rm repr}&\map\C^2\ox\C^2&\hbox{with}&\U(1)\ox\U(2)\cr
{\Phi\over M}&\mape v({\Phi\over M})_\al^{1,2},&&\hskip19mm\al=1,2\cr
\end{eq}and  
external Weyl and Pauli transmutators, parametrized with the
momenta as coset representatives (boost manifold, 2-sphere) 
\begin{eq}{rlcl}
s,\hat s:(\SL(\C^2)/\SU(2))_{\rm repr}
&\map \C^2\ox\C^2&\hbox{with}&\SU(2)\ox \SL(\C^2)\cr
{q\over m}&\mape 
s({q\over m})^{\dot A}_C,~\hat s({q\over m})^A_C,&&C=1,2;~~\dot A,A=1,2\cr
u:(\SU(2)/\SO(2))_{\rm repr}
&\map \C^2\ox\C^2&\hbox{with}&\SO(2)\ox \SU(2)\cr
{\rvec q\over |\rvec q|}&\mape 
u({\rvec q\over |\rvec q|})^\al_{1,2},&&\hskip19mm\al=1,2\cr
\end{eq}The operational triunits for the internal and external interaction-particle
transmutations are
\begin{eq}{lrcrll}
\hbox{Higgs }:&\U(1)&\x&[(\U(2)/\U(1)_+)_{\rm repr}\xs&\U(2)]\cr
\hbox{Weyl }:&\SU(2)&\x&[(\SL(\C^2)/\SU(2))_{\rm repr}\xs&\SL(\C^2)]\cr
\hbox{Pauli }:&\SO(2)&\x&[(\SU(2)/\SO(2))_{\rm repr}\xs&\SU(2)]\cr
\end{eq}

\end{document}